\documentclass[11pt]{article}

\usepackage{epsfig}

\begin{document}


\author{V. Kisel, G. Krylov, E. Ovsiyuk,
M. Amirfachrian,  V. Red'kov}
\title{
Wave functions of a particle with polarizability  \\ in the Coulomb
potential}

\maketitle

\begin{abstract}

 Quantum mechanical scalar particle with polarizability
 is considered in the presence of the Coulomb  field.
 Separation of   variables is performed with the use of Wigner
 $D$-functions, the radial system of 15 equations is reduced to a
 single second order differential equation, which among the
 Coulomb term includes an additional interaction term  of the form  $ \sigma  \alpha^{2}  / M^{2}  r^{4}$.
Various physical  regimes exist that is demonstrated by  examining
the behavior of the curves of generalized squared radial momentum
$P^{2}(r)$. Eigenstates of the equations can be constructed in
terms of double confluent Heun functions. Numerical analysis
proves the existence of the
 bound states in the  system; the lowest energy level and corresponding solution are calculated based on generalization
 of Ritz variational procedure.

\end{abstract}

\section{Separation of the variables}

A particle  with spin 0 and polarizability can be  described with
the use of tetrad  formalism by the following equation (the main
references to original papers concerning this model are given in
\cite{Book-1})
$$
[ \; \Gamma ^{\alpha }(x)\; ( \partial_{\alpha} \;  +  \;
B_{\alpha }(x) ) \; - m \;  ] \;\Psi  (x)  = 0 \; ,
$$
$$
\Gamma ^{\alpha }(x) = \Gamma ^{a} e ^{\alpha }_{(a)}(x) \; , \;
B_{\alpha }(x) = {1 \over 2}\; J^{ab} e ^{\beta }_{(a)}\nabla
_{\alpha }( e_{(b)\beta }) \; .
\eqno(1.1a)
$$

\noindent In Minkowski space and in spherical coordinates and
tetrad, eq.  $(1.1a)$ takes the form
$$
\left [\; \Gamma^{0} \partial_{0} \; + \;  \Gamma^{3}\partial_{r} +
\frac{\Gamma^{1} J^{31} + \Gamma^{2} J^{32} } {r}  \; + \;
\frac{1}{r}\; \Sigma_{\theta,\phi } \; - \; m \;\right  ] \; \Psi (x) = 0
\; , \eqno(1.1b)
$$
$$
\Sigma_{\theta,\phi} = \Gamma^{1} \; \partial_{\theta} \; + \;
\Gamma^{2} \; \frac{\partial_{\phi} +
\cos{\theta}J^{12}}{\sin{\theta}} \; . \eqno(1.1c)
$$

 General form for 15-component spherical wave function
  $\epsilon , j , m$  is  (more details see in \cite{Book-2}; the notation for Wigner functions is used;
  $D_{\sigma} = D^{j}_{-m, \sigma}(\phi, \theta, 0)$):
$$
C(x)  = e^{-i\epsilon t}   C (r)   D_{0} \; , \;
C_{0}(x) = e^{-i\epsilon t}    C_{0}(r)   D_{0}   \; , \;\;
\Phi_{0}(x) =    e^{-i\epsilon t} \Phi_{0} (x)  D_{0} \; ,
$$
$$
\vec{C} (x) =  e^{-i\epsilon t} \; \left | \begin{array}{l}
 C_{1}(r)  \; D_{-1}  \\
 C_{2}(3) \; D_{0}    \\
 C_{3}(r) \; D_{+1}      \end{array} \right | \; , \;
\vec{\Phi}(x)   = e^{-i\epsilon t}\; \left | \begin{array}{l}
\Phi_{1} (r)\; D_{-1}  \\
\Phi_{2}(r) \; D_{0}   \\
\Phi_{3}(r) \; D_{+1}  \end{array} \right | \; ,
$$
$$
\vec{E} (x)  = e^{-i\epsilon t} \; \left | \begin{array}{l}
E_{1}(r) \;  D_{-1} \\
E_{2}(r) \;  D_{0}   \\
E_{3}(r) \;  D_{+1}     \end{array} \right | \; , \;\; \vec{H} (x)
= e^{-i\epsilon t} \; \left | \begin{array}{l}
H_{1}(r)  \; D_{-1}  \\
H_{2}(r)  \; D_{0}   \\
H_{3}(r)  \; D_{+1}    \end{array} \right |\; .
  $$

\noindent After separation of  variables we obtain
the radial system
\cite{Book-2}
$$
-i \;  (\epsilon + {\alpha \over r} ) \; C_{0} \; -  \; ( {d \over
d r} \; + \; {2 \over r}) \; C_{2} \; - \; {\nu  \over r} \; (\;
C_{1} \; +  \; C_{3} \; )  =   m \; C \; ,
\eqno(1.2a)
$$
$$
-i \; (\epsilon + {\alpha \over r })  \; C \; - \;\sigma \;  (\;
{d \over d r} \; + \; {2 \over r} \;) \;  E_{2} \; - \; {\nu \;
\sigma \over r} \;   (\; E_{1} \; + \;  E_{3}\; ) = m \; C_{0} \;
,
$$
$$
i \; (\epsilon + {\alpha \over r} )\; \sigma \; E_{1} \; + \;
\sigma\;  ( \; {d \over dr} \; + \; {1 \over r }\; )\; H_{1} \; -
\; {\nu \over r} \;  C \; + \; {i \; \nu \; \sigma \over r} \;
H_{2} = m \; C_{1} \; ,
$$
$$
i \; (\epsilon + {\alpha \over r})  \; \sigma \; E_{2} \; + \; {d
\over dr}\; C \; - \; {i \; \nu \; \sigma \over r}\;
 (H_{1}\;-\;H_{3}) = m \;C_{2} \; ,
$$
$$
i \; ( \epsilon + {\alpha \over r} )\; \sigma \; E_{3} \; - \;
\sigma ( \; {d \over dr} \; +\; {1 \over r}\; )\; H_{3} \;
 -  \; {\nu \over r} \;  C \; - \; {i \; \nu \; \sigma \over r} \;  H_{2} = m \; C_{3} \; ,
\eqno(1.2b)
$$
$$
- i \; ( \epsilon + {\alpha \over r}) \; C = m \; \Phi _{0} \; ,
\qquad - {\nu \over r } \; C = m \;  \Phi_{1} \; ,
$$
$$
 {d \over dr }\; C = m \; \Phi_{2} \; ,
\;\;-\;  {\nu \over r}\;  C = m \; \Phi_{3} \; ,
\eqno(1.2c)
$$
$$
(\pm) \; [ \; - i \; ( \epsilon + { \alpha  \over }) \; \Phi_{1}
\; + \;  {\nu \over r} \; \Phi_{0} \; ] = m \; E_{1} \; ,
$$
$$
(\pm) \; [ \; - i\; ( \epsilon + { \alpha \over r} )\; \Phi_{2} \;
- \; {d \over dr}\; \Phi_{0} \; ] = m \; E_{2} \; ,
$$
$$
(\pm) \; [ \; - i \; (\epsilon + {\alpha \over r} ) \; \Phi_{3}
\;+ \;  {\nu \over r} \; \Phi_{0} \; ] = m \; E_{3} \; ,
$$
$$
(\pm)\; [ \;-i\; (\;  {d \over dr } \;+ \; {1 \over r}\; ) \;
\Phi_{1} - { i  \; \nu \over r }\;  \Phi_{2} \; \ = m \; H_{1}\; ,
$$
$$
(\pm)\;  {i \nu \over r}\;  (\; \Phi_{1} \; - \Phi_{3}\; ) \;  = m
\; H_{2} \; ,
$$
$$
(\pm)\; [ \; + i\;  (\; {d \over dr} \; + \; {1 \over r}\; ) \;
\Phi_{3}\; + \; { i \; \nu \over r}\;  \Phi_{2} \; ]  = m \;
H_{3}\; . \eqno(1.2d)
$$

\noindent
From $(1.2c)$ it follows that
$
\Phi _{3} = +   \Phi_{1}$. Then from  $(1.2d)$ we get
$
E_{3} = + E_{1}   , \; H_{3} = -  H_{1}  , \; H_{2} =
0$. Finally, from $(1.2b)$  it follows
$
C_{3} = + \; C_{1}$.
So, the collected  restrictions are
$$
\Phi _{3} = + \;  \Phi_{1} \; , \qquad C_{3} = + \; C_{1} \; ,
\qquad
E_{3} = + \;E_{1}  \; , \;\; H_{3} = - \; H_{1} \; , \;\; H_{2} =
0 \; . \eqno(1.3)
$$

Taking into account (1.3), the radial system reads
$$
-i \;  (\epsilon + {\alpha \over r} ) \; C_{0} \; -  \; ( {d \over
d r} \; + \; {2 \over r}) \; C_{2} \; - \; {\nu  \over r} \; 2
C_{1}   =   m \; C \; ,
\eqno(1.4a)
$$
$$
-i \; (\epsilon + {\alpha \over r })  \; C \; - \;\sigma \;  (\;
{d \over d r} \; + \; {2 \over r} \;) \;  E_{2} \; - \; {\nu \;
\sigma \over r} \;   \;  2  E_{1}  = m \; C_{0} \; ,
$$
$$
i \; (\epsilon + {\alpha \over r} )\; \sigma \; E_{1} \; + \;
\sigma\;  ( \; {d \over dr} \; + \; {1 \over r }\; )\; H_{1} \; -
\; {\nu \over r} \;  C \;  = m \; C_{1} \; ,
$$
$$
i \; (\epsilon + {\alpha \over r})  \; \sigma \; E_{2} \; + \; {d
\over dr}\; C \; - \; {i \; \nu \; \sigma \over r}\; 2H_{1}  = m
\;C_{2} \; ,
\eqno(1.4b)
$$
$$
- i \; ( \epsilon + {\alpha \over r}) \; C = m \; \Phi _{0} \; ,
\qquad - {\nu \over r } \; C = m \;  \Phi_{1} \; ,
$$
$$
 {d \over dr }\; C = m \; \Phi_{2} \; ,
\eqno(1.4c)
$$
$$
(\pm) \; [ \; - i \; ( \epsilon + { \alpha  \over }) \; \Phi_{1}
\; + \;  {\nu \over r} \; \Phi_{0} \; ] = m \; E_{1} \; ,
$$
$$
(\pm) \; [ \; - i\; ( \epsilon + { \alpha \over r} )\; \Phi_{2} \;
- \; {d \over dr}\; \Phi_{0} \; ] = m \; E_{2} \; ,
$$
$$
(\pm)\; [ \;-i\; (\;  {d \over dr } \;+ \; {1 \over r}\; ) \;
\Phi_{1} - { i  \; \nu \over r }\;  \Phi_{2} \; \ = m \; H_{1}\; .
\eqno(1.4d)
$$

\noindent Taking expressions for  $C_{i}$ according to  $(1.4c)$ , from
$(1.4d)$ one  gets
$$
E_{1} = 0     \; , \qquad  E_{2} = (\pm) \; ( - {i\; \alpha \over
m^{2} \; r^{2} }) \; C \; , \qquad H_{1} = 0 \; . \eqno(1.5)
$$

\noindent Now, from  $(1.4b)$ with the use of (1.5), one obtains
$$
m  C_{0} = -i  (\epsilon + {\alpha \over r})  \pm
{i\alpha \sigma \over m^{2} r^{2}}  {dC \over dr}
 \; ,\;
 m  C_{1} =  - {\nu \over r}  C \; ,
\qquad
m C_{2} = {dC \over dr}  \pm  {\alpha  \sigma \over m^{2}
 r^{2}} C\; . \eqno(1.6)
$$

\noindent Finally, we arrive at a second order differential equation
 $C(r)$ (changing  $m^{2}$  to$-M^{2}$,  and $2\nu^{2}$  to  $j(j+1)$):
$$
\left ( {d^{2} \over dr^{2} }  +  {2 \over r}\; {d \over dr}  + (\epsilon + {\alpha \over r } )^{2} - M^{2} \; - \;
{j(j+1) \over r^{2}} \; \pm \;{\sigma \; \alpha^{2} \over M^{2} \;
r^{4} } \right  )  C = 0 \; .
\eqno(1.7)
$$

\section{Qualitative  analysis of the radial equation }

With the use of notation  (the signs  $\pm$ can be included into
the parameter $\sigma$)
$$
\epsilon^{2} - M^{2} = - K^{2}, \qquad j(j+1)-  \alpha^{2}  = J^{2} \;
, \qquad
\sigma {\alpha^{2} \over M^{2}} = \Sigma^{2}
\eqno(2.1)
$$

\noindent  the main equation  (1.7) reads
$$
C (r) = {1 \over r} \; f (r) \; , \qquad
{d^{2} f \over dr^{2}}   +
 \left (       - K^{2}  + {2\epsilon \alpha \over r}
  -
 \frac{J^{2}}{r^{2}}
+ \frac{\Sigma^{2} }{r^{4}} \right ) f (r) = 0 \; .
\eqno(2.2)
$$

Let us examine the behavior of the squared radial momentum.
Near the origin and at infinity we have
$$
P^{2} (r \rightarrow 0 ) \sim  {\Sigma^{2} \over  r^{4}} \;, \qquad
P^{2} (r \rightarrow \infty  ) \sim  (\epsilon^{2} - M^{2} ) \;.
$$

\noindent To  describe classical turning points it is convenient to factorize the expression for  $P^{2} (r)$
$$
P^{2} (r)  = { (\epsilon^{2} - M^{2}) r^{4}   + 2 \epsilon \alpha  r^{3}
  -  J^{2} r^{2} + \Sigma^{2} \over r^{4} } =
$$
$$
  =   { (\epsilon^{2} - M^{2}) (r-r_{1})  (r-r_{2})  (r-r_{3})  (r-r_{4})  \over r^{4} } = 0 \; .
\eqno(2.3)
$$

\noindent The  roots  of the 4-th order polynomial obey
relations below
$$
-{2\epsilon \alpha \over \epsilon^{2} - M^{2} } =
r_{1} + r_{2} + r_{3} + r_{4} \; ,
\eqno(2.4a)
$$
$$
-{J^{2} \over \epsilon^{2} - M^{2} } = r_{1}r_{2}   + r_{1}r_{3} + r_{1} r_{4}
+ r_{2}r_{3} + r_{2} r_{4}  +  r_{3} r_{4}  \; ,
\eqno(2.4b)
$$
$$
0 = r_{1} r_{3} r_{4}  + r_{2}  r_{3} r_{4} +  r_{3}  r_{1} r_{2} + r_{4}  r_{1} r_{2} \; ,
\eqno(2.4c)
$$
$$
{\Sigma^{2} \over \epsilon^{2} - M^{2} } = r_{1} r_{2} r_{3} r_{4} \; .
\eqno(2.4d)
$$

First,  let us consider the case of the  {\bf bound states}.
There exist two different possibilities depending on
the sign of $\Sigma^{2}$.
$$
I \qquad
\epsilon^{2} - M^{2} < 0 \; , \qquad  \Sigma^{2} = \sigma (\alpha^{2} /  M^{2} )  < 0 \; ;
\eqno(2.5a)
$$

\noindent so, from  (2.4) we conclude that  two roots  can be positive
and two negative (or complex  and conjugate to each other)
$$
r_{1} < 0 \; ,  \qquad  r_{2} <  0 \; , \qquad
r_{3} > 0\;  ,   \qquad r_{4}  > 0 \;  .
\eqno(2.5b)
$$

\noindent
that can be illustrated by Fig.~1.

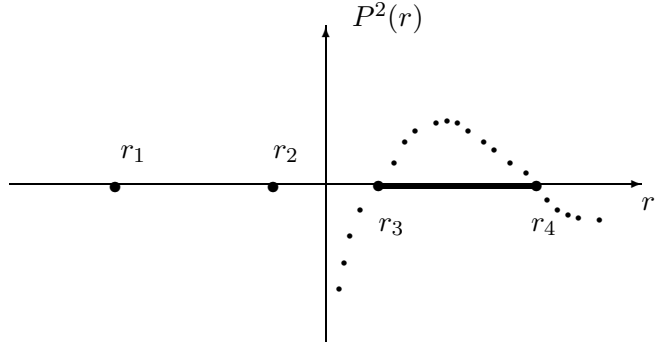
\begin{figure}
 \unitlength=0.7 mm
\begin{picture}(160,100)(-120,-60)
\special{em:linewidth 0.4pt} \linethickness{0.4pt}

\put(-60,0){\vector(+1,0){120}}     \put(+60,-5){$r $}
\put(0,-30,0){\vector(0,+1){60}}   \put(+5,+30){$P^{2}(r) $}

\put(+40,-0.5){\circle*{2}}  \put(+39,-9){$r_{4}$}
\put(+10,-0.5){\circle*{2}}  \put(+10,-9){$r_{3}$}

\put(10,-0.5){\line(+1,0){30} }
\put(10,-0.8){\line(+1,0){30} }
\put(10,-0.3){\line(+1,0){30} }

\put(-40,-0.5){\circle*{2}}  \put(-39,+5){$r_{1}$}
\put(-10,-0.5){\circle*{2}}  \put(-10,+5){$r_{2}$}

\put(+2.5,-20){\circle*{1}}
\put(+3.5,-15){\circle*{1}}
\put(+4.6,-10){\circle*{1}}
\put(+6.5,-5){\circle*{1}}
\put(+13,+4){\circle*{1}}
\put(+15,+8){\circle*{1}}
\put(+17,+10){\circle*{1}}
\put(+21,+11.5){\circle*{1}}
\put(+23,+12){\circle*{1}}
\put(+25,+11.5){\circle*{1}}
\put(+27,+10){\circle*{1}}
\put(+30,+8){\circle*{1}}
\put(+32,+6.5){\circle*{1}}
\put(+35,+4){\circle*{1}}
\put(+38,+2){\circle*{1}}

\put(+42,-3){\circle*{1}}
\put(+44,-5){\circle*{1}}
\put(+46,-6){\circle*{1}}
\put(+48,-6.5){\circle*{1}}
\put(+52,-6.8){\circle*{1}}
\end{picture}
\vspace{-20mm}
\caption{Finite classical motion:  $r \in [ r_{3}, r_{4}]$}

\end{figure}

With the use of  (2.4a)  and  (2.4b) one expresses the roots
$r_{1}, r_{2}$  through two classical turning points   $r_{3}, r_{4}$
$$
r_{1} + r_{2} =  {2\epsilon \alpha \over   M^{2} - \epsilon^{2}  }  -  r_{3} - r_{4}
  \; , \qquad  r_{1} r_{2}  =
{\Sigma^{2} \over \epsilon^{2} - M^{2} }{1 \over  r_{3} r_{4} } \; \Longrightarrow
$$
$$
r_{1} ={1\over 2} \, \left[
- \left(r_{3}+r_{4} -{2 \epsilon \alpha\over M^{2}- \epsilon^{2}}\right)
-   \sqrt{   \left(r_{3}+r_{4}-  {2\epsilon \alpha\over M^{2}- \epsilon^{2}}\right)^{2}
+{4\Sigma^{2}\over(M^{2}- \epsilon^{2})r_{3}r_{4}}}\;\right]\,,
$$
$$
r_{2}= {1\over 2}\,\left[-\left(r_{3}+r_{4}-{2 \epsilon \alpha\over M^{2}-\epsilon^{2}}\right)
+\sqrt{\left(r_{3}+r_{4} -{2\epsilon \alpha\over M^{2} - \epsilon^{2}}\right)^{2}
+
{4\Sigma^{2}\over(M^{2}- \epsilon^{2})r_{3}r_{4}}}\;\right]\,;
$$
$$
\eqno(2.5c)
$$

\noindent
To obtain two positive and two negative roots we  require  two conditions
$$
-(r_{1} +r_{2}) = r_{3} + r_{4} - {2\epsilon \alpha \over M^{2} - \epsilon^{2}} > 0 \; ,
$$
$$
 \left(r_{3}+r_{4}-  {2\epsilon \alpha\over M^{2}- \epsilon^{2}}\right)^{2}
+{4\Sigma^{2}\over(M^{2}- \epsilon^{2})r_{3}r_{4}} > 0 \; .
\eqno(2.5d)
$$

\noindent
The case with two positive and two complex   roots (with two negative real parts)
  is realized if two following   conditions are imposed
$$
-(r_{1} +r_{2}) = r_{3} + r_{4} - {2\epsilon \alpha \over M^{2} - \epsilon^{2}} > 0 \; ,
$$
$$
 \left(r_{3}+r_{4}-  {2\epsilon \alpha\over M^{2}- \epsilon^{2}}\right)^{2}
+{4\Sigma^{2}\over(M^{2}- \epsilon^{2})r_{3}r_{4}} <  0 \;.
\eqno(2.5e)
$$

It should  be noted that another  possibility exists for the bound states
at positive values for $\Sigma^{2}$:
$$
II \qquad
\epsilon^{2} - M^{2} < 0 \; , \qquad
\Sigma^{2} = \sigma {\alpha^{2} \over M^{2}} > 0 \; ;
\eqno(2.6a)
$$

\noindent from (2.4) we conclude that three  roots can be positive and  one negative
$$
r_{1} < 0 \; ,  \qquad  r_{2} >   0 \; , \qquad
r_{3} > 0\;  ,   \qquad r_{4}  > 0 \;  ,
$$
$$
r_{1}  = -    { r_{2}  r_{3} r_{4}  \over  r_{3} r_{4}   +  r_{3}   r_{2} + r_{4}   r_{2} }\; ,
\;
  r_{1}  = (   r_{2} + r_{3} + r_{4}  - {2\epsilon \alpha \over  M^{2} - \epsilon^{2} } ) < 0 \; .
\eqno(2.6b)
$$

\noindent
This case is illustrated in Fig.  2.

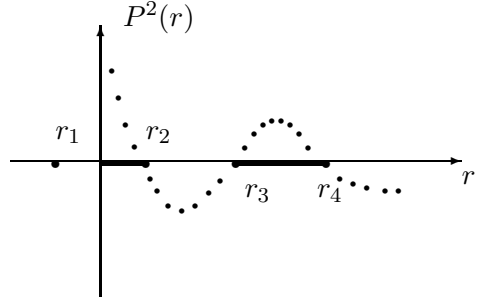
\begin{figure}[hbt]
 \unitlength=0.6mm
\begin{picture}(160,100)(-120,-60)
\special{em:linewidth 0.4pt} \linethickness{0.4pt}

\put(-20,0){\vector(+1,0){100}}     \put(+80,-5){$r$}
\put(0,-30,0){\vector(0,+1){60}}   \put(+5,+30){$P^{2}(r) $}

\put(-10,-0.5){\circle*{2}}  \put(-10,+5){$r_{1}$}

\put(+10,-0.5){\circle*{2}}  \put(+10,+5){$r_{2}$}

\put(+30,-0.5){\circle*{2}}  \put(+32,-8){$r_{3}$}

\put(+50,-0.5){\circle*{2}}  \put(+48,-8){$r_{4}$}

\put(10,-0.5){\line(-1,0){10} }
\put(10,-0.8){\line(-1,0){10} }
\put(10,-0.3){\line(-1,0){10} }

\put(30,-0.5){\line(+1,0){20} }
\put(30,-0.8){\line(+1,0){20} }
\put(30,-0.3){\line(+1,0){20} }

\put(+2.5,+20){\circle*{1}}
\put(+4,+14){\circle*{1}}
\put(+5.5,+8){\circle*{1}}
\put(+7.5,+3.2){\circle*{1}}

\put(+11,-4){\circle*{1}}
\put(+12.5,-6.5){\circle*{1}}

\put(+15,-10){\circle*{1}}
\put(+18,-11){\circle*{1}}
\put(+21,-10){\circle*{1}}
\put(+24,-7){\circle*{1}}
\put(+26.5,-4.5){\circle*{1}}

\put(+32, +3){\circle*{1}}
\put(+34, +6){\circle*{1}}
\put(+36, +8){\circle*{1}}
\put(+38, +9){\circle*{1}}
\put(+40, +9){\circle*{1}}
\put(+42, +8){\circle*{1}}
\put(+44, +6){\circle*{1}}
\put(+46, +3){\circle*{1}}

\put(+53, -4){\circle*{1}}
\put(+56, -5){\circle*{1}}
\put(+59, -6){\circle*{1}}
\put(+63, -6.5){\circle*{1}}
\put(+66, -6.6){\circle*{1}}

\end{picture}
\vspace{-15mm}

\caption{Finite classical motion: $r \in [0, r_{2}], \; r \in [r_{3}, r_{4}] $}

\end{figure}

\noindent
The formulas for two roots  $r_{1} $ and $r_{2}$ in terms of  $r_{3}, r_{4}$ are
$$
r_{1} ={1\over 2} \, \left[
- \left(r_{3}+r_{4} -{2 \epsilon \alpha\over M^{2}- \epsilon^{2}}\right)
-   \sqrt{   \left(r_{3}+r_{4}-  {2\epsilon \alpha\over M^{2}- \epsilon^{2}}\right)^{2}
+{4\Sigma^{2}\over(M^{2}- \epsilon^{2})r_{3}r_{4}}}\;\right]\,,
$$
$$
r_{2}= {1\over 2}\,\left[-\left(r_{3}+r_{4}-{2 \epsilon \alpha\over M^{2}-\epsilon^{2}}\right)
+\sqrt{\left(r_{3}+r_{4} -{2\epsilon \alpha\over M^{2} - \epsilon^{2}}\right)^{2}
+
{4\Sigma^{2}\over(M^{2}- \epsilon^{2})r_{3}r_{4}}}\;\right]\,;
$$
$$
\eqno(2.6c)
$$

\noindent
but now the parameter is negative $\Sigma^{2}>0$, and correspondingly
the root $r_{2}$  is positive.

The situation when  one root is positive ant three are negative is also possible
$$
r_{1} < 0 \; ,  \qquad  r_{2} >   0 \; , \qquad
r_{3} < 0\;  ,   \qquad r_{4}  > 0 \;  ,
$$
$$
r_{4}  = -    { r_{1}  r_{2} r_{3}  \over  r_{1} r_{2}   +  r_{1}   r_{3} + r_{2}   r_{3} }\; ;
\eqno(2.6d)
$$

\noindent
that is illustrated in Fig. 3.

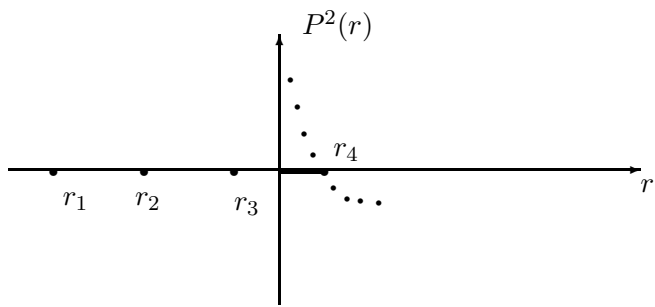
\begin{figure}[hbt]
 \unitlength=0.6 mm
\begin{picture}(160,100)(-110,-60)
\special{em:linewidth 0.4pt} \linethickness{0.4pt}

\put(-60,0){\vector(+1,0){140}}     \put(+80,-5){$r$}
\put(0,-30,0){\vector(0,+1){60}}   \put(+5,+30){$P^{2}(r) $}

\put(-10,-0.5){\circle*{2}}  \put(-10,-9){$r_{3}$}

\put(+10,-0.5){\circle*{2}}  \put(+12,+3){$r_{4}$}

\put(-30,-0.5){\circle*{2}}  \put(-32,-8){$r_{2}$}

\put(-50,-0.5){\circle*{2}}  \put(-48,-8){$r_{1}$}

\put(10,-0.5){\line(-1,0){10} }
\put(10,-0.8){\line(-1,0){10} }
\put(10,-0.3){\line(-1,0){10} }

\put(+2.5,+20){\circle*{1}}
\put(+4,+14){\circle*{1}}
\put(+5.5,+8){\circle*{1}}
\put(+7.5,+3.2){\circle*{1}}

\put(+12,-4){\circle*{1}}
\put(+15,-6.5){\circle*{1}}
\put(+18,-7){\circle*{1}}
\put(+22,-7.3){\circle*{1}}

\end{picture}
\vspace{-10mm}
\caption{Finite classical motion:  $r \in [0, r_{4}]$}
\end{figure}

Now let us consider possible infinite motions.
The first possibility is
$$
III \qquad
\epsilon^{2} - M^{2} > 0 \; , \qquad  \Sigma^{2} = \sigma (\alpha^{2} /  M^{2} )  < 0 \; ;
\eqno(2.7a)
$$
$$
r_{1} < 0 \; ,  \qquad  r_{2} >  0 \; , \qquad
r_{3} > 0\;  ,   \qquad r_{4}  > 0 \;  .
\eqno(2.7b)
$$

\noindent that  is illustrated in Fig.~4.


\begin{figure}[hbt]

 \unitlength=0.6 mm
\begin{picture}(160,100)(-110,-60)
\special{em:linewidth 0.4pt} \linethickness{0.4pt}

\put(-60,0){\vector(+1,0){150}}     \put(+90,-5){$r $}
\put(0,-30,0){\vector(0,+1){60}}   \put(+5,+30){$P^{2}(r) $}

\put(+40,-0.5){\circle*{2}}  \put(+39,-9){$r_{3}$}
\put(+10,-0.5){\circle*{2}}  \put(+10,-9){$r_{2}$}

\put(10,-0.5){\line(+1,0){30} }
\put(10,-0.8){\line(+1,0){30} }
\put(10,-0.3){\line(+1,0){30} }

\put(-40,-0.5){\circle*{2}}  \put(-39,+5){$r_{1}$}
\put(+60,-0.5){\circle*{2}}  \put(+60,-9){$r_{4}$}

\put(+2.5,-20){\circle*{1}}
\put(+3.5,-15){\circle*{1}}
\put(+4.6,-10){\circle*{1}}
\put(+6.5,-5){\circle*{1}}
\put(+13,+4){\circle*{1}}
\put(+15,+8){\circle*{1}}
\put(+17,+10){\circle*{1}}
\put(+21,+11.5){\circle*{1}}
\put(+23,+12){\circle*{1}}
\put(+25,+11.5){\circle*{1}}
\put(+27,+10){\circle*{1}}
\put(+30,+8){\circle*{1}}
\put(+32,+6.5){\circle*{1}}
\put(+35,+4){\circle*{1}}
\put(+38,+2){\circle*{1}}

\put(+42,-3){\circle*{1}}
\put(+44,-5){\circle*{1}}
\put(+46,-6){\circle*{1}}
\put(+48,-6.5){\circle*{1}}
\put(+51,-6.5){\circle*{1}}
\put(+54,-4.5){\circle*{1}}
\put(+57,-2.5){\circle*{1}}

\put(+62,+2){\circle*{1}}
\put(+64,+3.5){\circle*{1}}
\put(+67,+4){\circle*{1}}
\put(+69,+4.5){\circle*{1}}

\put(+60,-0,2){\line(+1,0){20}}
\put(+60,+0,2){\line(+1,0){20}}
\put(+60,-0,4){\line(+1,0){20}}
\put(+60,+0,4){\line(+1,0){20}}

\end{picture}

\vspace{-20mm}

\caption{Infinite classical motion: $r \in [r_{2}, r_{3}], r \in [r_{4}, + \infty )$}

\end{figure}
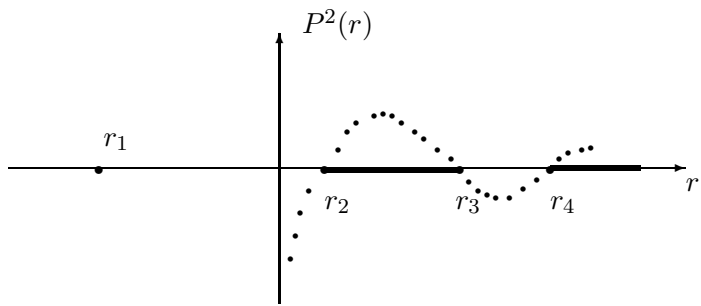

Besides, another case  can be realized
if three roots are negative and one is positive
$$
r_{1} < 0 \; ,  \qquad  r_{2} <  0 \; , \qquad
r_{3} < 0\;  ,   \qquad r_{4}  > 0 \;  ;
\eqno(2.7c)
$$

\noindent that is illustrated in Fig.~5.

\begin{figure}[bt]
 \unitlength=0.6mm
\begin{picture}(160,100)(-110,-60)
\special{em:linewidth 0.4pt} \linethickness{0.4pt}

\put(-60,0){\vector(+1,0){120}}     \put(+60,-5){$r $}
\put(0,-30,0){\vector(0,+1){60}}   \put(+5,+30){$P^{2}(r) $}

\put(-40,-0.5){\circle*{2}}  \put(-40,+5){$r_{1}$}
\put(-20,-0.5){\circle*{2}}  \put(-20,+5){$r_{2}$}
\put(-10, -0.5){\circle*{2}}  \put(-10,+5){$r_{3}$}

\put(+10,-0.5){\circle*{2}}  \put(+10,-9){$r_{4}$}

\put(10,-0.6){\line(+1,0){40} }
\put(10,-0.8){\line(+1,0){40} }
\put(10,-0.3){\line(+1,0){40} }
\put(10,+0.3){\line(+1,0){40} }

\put(+2.5,-20){\circle*{1}}
\put(+3.5,-15){\circle*{1}}
\put(+4.6,-10){\circle*{1}}
\put(+6.5,-5){\circle*{1}}
\put(+13,+4){\circle*{1}}
\put(+16,+7){\circle*{1}}
\put(+19,+8){\circle*{1}}
\put(+23,+8.5){\circle*{1}}
\put(+27,+8.8){\circle*{1}}

\end{picture}

\vspace{-20mm}
\caption{Infinite classical motion:  $r \in [r_{4}, + \infty )$}
\end{figure}
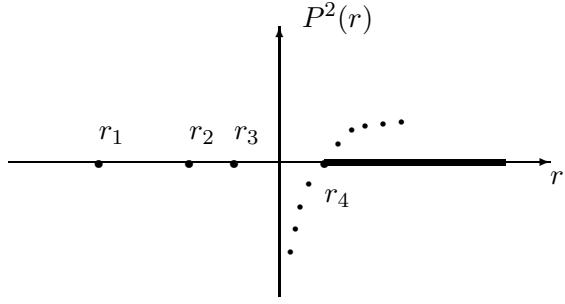

Finally, there exists one more case
$$
IV \qquad
\epsilon^{2} - M^{2} > 0 \; , \qquad  \Sigma^{2} = \sigma (\alpha^{2} /  M^{2} )  > 0 \; ;
\eqno(2.8a)
$$

\noindent when two root are positive and two roots are negative
$$
r_{1} < 0 \; ,  \qquad  r_{2} <  0 \; , \qquad
r_{3} > 0\;  ,   \qquad r_{4}  > 0 \;  ;
\eqno(2.8b)
$$

\noindent
that is illustrated in Fig.~6.

\begin{figure}[hbt]

 \unitlength=0.6 mm
\begin{picture}(160,100)(-90,-60)
\special{em:linewidth 0.4pt} \linethickness{0.4pt}

\put(-20,0){\vector(+1,0){80}}     \put(+60,-5){$r $}
\put(0,-30,0){\vector(0,+1){60}}   \put(+5,+30){$P^{2}(r) $}

\put(-25,-0.5){\circle*{2}}  \put(-28,+5){$r_{1}$}
\put(-15,-0.5){\circle*{2}}  \put(-15,+5){$r_{2}$}

\put(+10,-0.5){\circle*{2}}  \put(+5,-8){$r_{3}$}
\put(+30,-0.5){\circle*{2}}  \put(+30,-9){$r_{4}$}

\put(10,-0.5){\line(-1,0){10} }
\put(10,-0.8){\line(-1,0){10} }
\put(10,-0.3){\line(-1,0){10} }

\put(30,-0.5){\line(+1,0){28} }
\put(30,-0.8){\line(+1,0){28} }
\put(30,-0.3){\line(+1,0){28} }

\put(+2.5,+20){\circle*{1}}
\put(+4,+14){\circle*{1}}
\put(+5.5,+8){\circle*{1}}
\put(+7.5,+3.2){\circle*{1}}

\put(+11,-4){\circle*{1}}
\put(+12.5,-6.5){\circle*{1}}

\put(+15,-10){\circle*{1}}
\put(+18,-11){\circle*{1}}
\put(+21,-10){\circle*{1}}
\put(+24,-7){\circle*{1}}
\put(+26.5,-4.5){\circle*{1}}

\put(+32, +3){\circle*{1}}
\put(+34, +6){\circle*{1}}
\put(+36, +8){\circle*{1}}
\put(+38, +9){\circle*{1}}

\end{picture}

\vspace{-20mm}

\caption{ Fig.  6.   Infinite classical motion:  $R \in [0, r_{2} ], \; r \in [r_{4}, + \infty )$}

\end{figure}
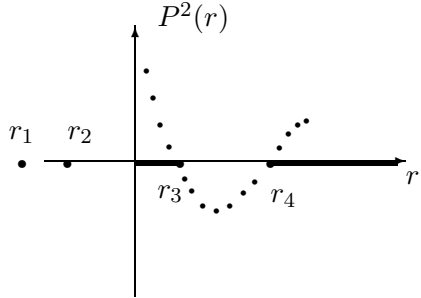

\section{ Analytical treatment of the problem  }

Let us turn back to eq.   (2.2) and introduce a new variable
$$
x={i(-K^{2}\Sigma^{2})^{1/4}r+\Sigma\over i(-K^{2}\Sigma^{2})^{1/4}r-\Sigma}\,,
\qquad
r=-{i\Sigma\over (-K^{2}\Sigma^{2})^{1/4}} \; { (x+1) \over  (x-1) }\,.
\eqno(3.1a)
$$

\noindent
it is readily verified the main physical singularities
 $r = 0$  and $r= \infty$ in the new  variable look as
 $$
r=\infty  \; \Longrightarrow \; x= + 1\,, \qquad
r=0  \; \Longrightarrow  \; x=-1\,.
\eqno(3.1b)
$$

\noindent
In general, the variable   $x$ is complex-valued
$$
x(r) = {iA r + \Sigma
\over iA r - \Sigma } \;, \qquad (-K^{2}\Sigma^{2})^{1/4} = [ (\epsilon^{2} - M^{2}) \Sigma^{2} ]^{1/4}= A \; .
\eqno(3.1c)
$$

\noindent
There exist four different combinations of signs
$$
(\epsilon^{2} - M^{2}, \;  \Sigma^{2} )  \;\;\; \Longrightarrow \;\;\;
\left \{  \begin{array}{cc}
(+, +) & \qquad (+,- )\\
(-, +) & \qquad (- , -)
\end{array}
 \right \} \; .
\eqno(3.2a)
 $$

Below we will consider only the case of bound states when
 $\epsilon^{2} - M^{2} < 0$, which corresponds to variants  $(-, +)$  and $(- , -)$.
If  $\Sigma^{2}$   (the case $(-, +))$, the variable  $x$  is complex-valued
$$
x(r) = {iA r + \Sigma
\over iA r - \Sigma } = - {  \Sigma^{2} - A^{2} r^{2}  \over   \Sigma^{2}  + A^{2} r^{2} }
- i {2A\Sigma r  \over   \Sigma^{2}  + A^{2} r^{2} } = e^{i \varphi(r)} \;, \qquad \mid x(r) \mid = 1 \; ;
\eqno(3.2b)
$$

\noindent that is illustrated in Fig.~7.

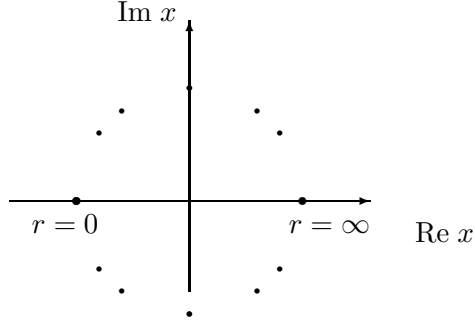
\begin{figure}[hbt]
 \unitlength=0.6 mm
\begin{picture}(160,100)(-120,-60)
\special{em:linewidth 0.4pt} \linethickness{0.4pt}

\put(-40,0){\vector(+1,0){80}}    \put(+50,-9){$\mbox{Re}\; x$ }

\put(0,-20){\vector(0,+1){60}}      \put(-16,+40){$\mbox{Im}\; x$}

\put(+25,0){\circle*{2}}
\put(+20,+15){\circle*{1}}
\put(+15,+20){\circle*{1}}
\put(0,+25){\circle*{1}}
\put(-25,0){\circle*{2}}
\put(-20,+15){\circle*{1}}
\put(-15,+20){\circle*{1}}
\put(0,+25){\circle*{1}}

\put(+20,-15){\circle*{1}}
\put(+15,-20){\circle*{1}}
\put(0,-25){\circle*{1}}
\put(-25,0){\circle*{2}}
\put(-20,-15){\circle*{1}}
\put(-15,-20){\circle*{1}}
\put(0,-25){\circle*{1}}

\put(+22,-7){$r=\infty $}   \put(-35,-7){$r= 0 $}

\end{picture}

\vspace{-20mm}

\caption{Variable $x$ on complex plane}

\end{figure}

If  $\Sigma^{2} = - b^{2} < 0 $  (the case  $(-, -))$, the variable  $x$  is real-valued
$$
x(r) = {A r + b
\over A r - b } \; , \qquad x \in [\; -1 , \; + 1 \; ]\; .
\eqno(3.2c)
$$

With the variable $x$, the differential equation (2.2) takes the form
$$
{d^{2}\over dx^{2}} \Omega +{2\over x-1}\,{d\over dx} \Omega +
\left(\; {4K^{2}\Sigma^{2}\over \sqrt{-K^{2}\Sigma^{2}}
(x+1)^{4}}+{4K^{2}\Sigma^{2}\over \sqrt{-K^{2}\Sigma^{2}}(x-1)^{4}}-\right.
$$
$$
\left.
-
{4\,i\,\epsilon \,\alpha \,\Sigma\over(-K^{2}\Sigma^{2})^{1/4}(x-1)^{3}}
-{J^{2}\over(x+1)^{2}}+{2i\epsilon\alpha\Sigma-J^{2}(-K^{2}\Sigma^{2})^{1/4}\over
(-K^{2}\Sigma^{2})^{1/4}(x-1)^{2}}+
\right.
$$
$$
\left. + {i\epsilon\alpha\Sigma-J^{2}(-K^{2}\Sigma^{2})^{1/4}\over(-K^{2}\Sigma^{2})^{1/4}(x+1)}+{-i\epsilon\alpha\Sigma+
J^{2}(-K^{2}\Sigma^{2})^{1/4}\over(-K^{2}\Sigma^{2})^{1/4}(x-1)}\right) f (x) =0\,.
\eqno(3.3)
$$

\noindent
With the use of the notation
$$
(-K^{2}\Sigma^{2})^{1/4} = A\;, \qquad -K^{2}\Sigma^{2} = A^{4}\; , \qquad
A^{2 } = \sqrt{-K^{2}\Sigma^{2}}  =  \pm i K \Sigma \; ,
$$

\noindent
 eq.  (3.3) reads
 $$
{d^{2}\over dx^{2}} f +{2\over x-1}\,
{d\over dx} f +
\left( - {4A^{2} \over (x+1)^{4}} -{4A^{2} \over
(x-1)^{4}}-{4\,i\,\epsilon \,\alpha \,\Sigma\over A(x-1)^{3}}-\right.
$$
$$
\left. -{J^{2}\over(x+1)^{2}}+{2i\epsilon\alpha\Sigma-J^{2}A \over A(x-1)^{2}}+{i\epsilon\alpha\Sigma-J^{2}A\over A(x+1)}+{-i\epsilon\alpha\Sigma+
J^{2}A\over A(x-1)}\right) f (x) =0\,.
\eqno(3.4)
$$

\noindent
Applying the substitution
$$
f (x) =(x+1)^{B}(x-1)^{C}\exp\left({Dx\over (x+1)(x-1)}\right) F(x)
\eqno(3.5)
$$

\noindent we arrive at the equation for $F$
$$
{d^{2}F\over dx^{2}}+
\left ({2B\over x+1}+{2C+2\over x-1}-{D\over (x+1)^{2}}-{D\over (x-1)^{2}}\right )
{df\over dx} +
$$
$$
+
\left ({D^{2}- 16 A^{2}\over 4 (x+1)^{4}}
+{D^{2} -16 A^{2} \over 4(x-1)^{4}}-\right.
$$
$$
-{D(B-1)\over (x+1)^{3}}-{CD\,A+4i\epsilon \alpha \Sigma\over A (x-1)^{3}}+{D^{2}+8B^{2}-8B+4CD+4D-8J^{2}\over 8\,(x+1)^{2}}+
$$
$$
+{D^{2}A+8CA-4BD\,A+8C^{2}A+16i \epsilon \alpha \Sigma-8J^{2}A\over 8\,A(x-1)^{2}}+
$$
$$
+{D^{2}A+2CD\,A-2BD\,A-8BC\,A+2D\,A-8B\,A+8 i \epsilon \alpha \Sigma-8J^{2}A\over 8A(x+1)}+
$$
$$
\left.+{-D^{2}A-2CD\,A+2BD\,A+8BC\,A-2D\,A+8B\,A-8 i \epsilon \alpha \Sigma+8J^{2}A\over 8A(x-1)}\right ) F=0\,.
$$
$$
\eqno(3.6)
$$

When   $B,\;C,\;D$ are given as (see (3.5))
$$
B={1\over 2}\,\qquad C=-{1\over 2}\,,\qquad D=\pm\,{4iK\Sigma\over A} = \pm 4  A \,,
$$
$$
f=\sqrt{ {x+1 \over x-1}} \; \exp\left({Dx\over (x+1)(x-1)}\right) \; F(x)
\eqno(3.7)
$$

\noindent eq.  (3.6) becomes more simple
$$
{d^{2}F\over dx^{2}}+\left[{1\over x+1}+{1\over x-1}-{D\over (x+1)^{2}}-{D\over (x-1)^{2}}\right]\,{dF\over dx}+
$$
$$
+{1 \over 2A(x+1)^{3}(x-1)^{3}} \;
\left [ \; (D^{2}A-8J^{2}A-2A-16 i \epsilon \alpha \Sigma)x^{2}+ \right.
$$
$$
\left. + (8DA-32 i \epsilon \alpha \Sigma)x+2A-16 i \epsilon
 \alpha \Sigma-D^{2}A+8J^{2}A \; \right  ]\;F=0\,.
\eqno(3.8)
$$

\noindent It coincides with  the double confluent Heun equation
 \cite{Ronveaux-Arscott, Slavyanov-Lay}
 for  $H(\mu,\,\beta,\,\gamma,\,\delta,\,z)$:
$$
{d^{2}H \over dx^{2}}+\left ({1\over x+1}+{1\over x-1}-{\mu\over 2(x+1)^{2}}-{\mu\over 2(x-1)^{2}}\right) {dH\over dx}+
$$
$$
{\beta x^{2}+(\gamma+2 \mu)x+\delta\over (x+1)^{3}(x-1)^{3}}\,H=0\,
\eqno(3.9a)
$$

\noindent with parameters
$$
\mu=2D = \pm 8  A \,, \qquad
\gamma=-{16 i \epsilon \alpha \Sigma\over A} \; ,
$$
$$
\beta= - 1 - 4J^{2}   +8 A^{2} -8 i \epsilon \alpha {\Sigma \over A }  \,,
$$
$$
\; \delta= + 1 +4J^{2}  -8A^{2} -8 i \epsilon \alpha {\Sigma \over A}  \,;
\eqno(3.9b)
$$

\noindent
with additional constrain
$
\beta + \delta =  \gamma$  .

\section{Numerical simulations }

To  possibility for quantum mechanical  bound states with energy $\epsilon \i [-M,+M]$
there must corresponds in classical description a finite region for classical motion.
Let us examine the condition at which the 4-nd order polynomial (this analysis will enable us to
find  left limiting  boundary  $V$ for  possible  quantum energy levels,
$\epsilon \geq  \epsilon_{0}  $):
$$
\Pi (r) = { (\epsilon^2-M^2)r^4 +2\epsilon \alpha r^3 -j(j+1) r^2 +\sigma
\alpha^2/M^2 \over \epsilon^2-M^2 } =0
\eqno(4.1)
$$

\noindent just starts to have a double root  $r_0$  at positive real axis
$$
\Pi (r) = (r-r_0)^2(r-a+i b)(r-a-i b)=0 \; .
\eqno(4.2)
$$
This is just the bifurcation value of $e$.

It is convenient to measure $\epsilon$  in unit of $M$ introducing $\epsilon= e M$.
From comparing (4.1) and (4.2) we get the system of algebraic equations
for $r_0,a,b,e$
$$
- (a^2 + b^2) r_0^2 +  { \alpha^2 \sigma  \over (-1 + e^2) M^4 }=0 \; ,
$$
$$
2 r_0 (a^2 + b^2 + a r_0)=0 \; ,
$$
$$-a^2 - b^2 - {j (1 + j) \over (-1 + e^2) M^2 } - 4 a r_0 - r_0^2=0 \; ,
$$
$$
2 (a + r_0) - { 2 e \alpha \over M - e^2 M }=0\; . \eqno(4.3)
$$

From whence it follows
$$
a =  { M (r_0 - e^2 r_0) - e \alpha \over
(-1 + e^2) M} \; ,
$$
$$
b^2 = - { (e \alpha ((-1 + e^2) M r_0 + e \alpha) \over
(-1 + e^2)^2 M^2 } \; ,
$$
$$
r_0 = - {
  3 e \alpha + \sqrt{8 (-1 + e^2) j (1 + j) + 9 e^2 \alpha^2 } \over
 4 (-1 + e^2) M} \; ,
$$
and we arrive at a rather complicated equation for $e$
$$
-16 (-1 + e^2)^2 j^3 - 8 (-1 + e^2)^2 j^4 - 27 e^4 \alpha^4 -
$$
$$
 9 e^3 \alpha^3 \sqrt{8 (-1 + e^2) j (1 + j) + 9 e^2 \alpha^2} -
 $$
 $$
 -4 e (-1 + e^2) j \alpha (9 e \alpha +
    2 \sqrt{8 (-1 + e^2) j (1 + j) + 9 e^2 \alpha^2}) -
    $$
    $$
 4 (-1 + e^2) j^2 (-2 + e^2 (2 + 9 \alpha^2) +
 $$
 $$
    2 e \alpha \sqrt{8 (-1 + e^2) j (1 + j) + 9 e^2 \alpha^2}) +
 32 (-1 + e^2)^3 \alpha^2 \sigma =0
 \eqno(4.4)
 $$

There exists additional constraint (condition for real-valuedness of $r_0$(
$$
|e| > \sqrt{\frac{8 j + 8 j^2}{8 j + 8 j^2 + 9 \alpha^2}} \; .
 \eqno(4.5)
$$

Numerically one  easily finds the  value of $e_{\mbox{min}}$ which is a lower boundary  for the existence
of bound states

At $j=0,\sigma=-1,\alpha=1,M=1$ we obtain $e_{\mbox{min}}=0.614659$, the  plot of curve
 is illustrated in Fig.~8.

\begin{figure*}[hbt]
\includegraphics[width=12cm]{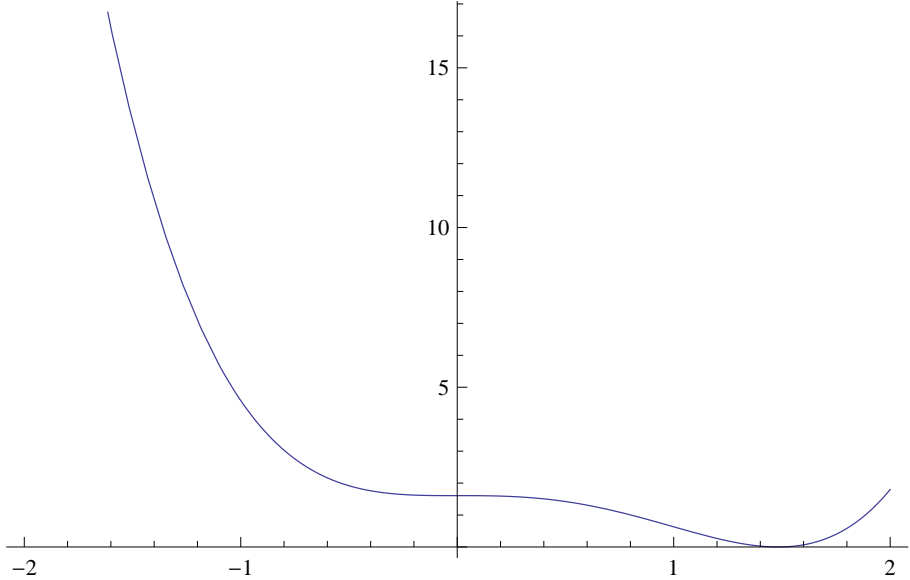}
 \caption { $\Pi (r$) at $j=0,\sigma=-1,\alpha=1,M=1$}
\end{figure*}

As one can find by numerical simulations at  $\sigma>$ possible values for $e$
lay outside  the interval [-1,1] and lead to $r_0<0$.

After obtaining information of regions of possible bound state energy $e$ for the system with
small quantum number $j,\ldots$ we can start to construct numerical solution for the lowest eigenstate.
To this goal we extend the known Ritz variational approach
\cite{Landau-3}, performing simulations for simplicity at fixed values
$M=1, \sigma=-1, j=0$.

First we find the asymptotic behavior of the eq. (1.7) at the origin.
Selecting the most singular terms we get
 $$
  f''(r) -\alpha^2 f(r)/r^4=0
 $$
with the leading term in asymptotic solution of the form
$r\ll 1$, $f(r)=1/(2\alpha)
\exp(-\alpha/r) r$.
In a similar way at infinity we introduce the variable
$u=1/r$, rewriting the equation and selecting most singular terms only we get
$$
y''(u) -y(u)/u^4(1-\epsilon^2)=0
 \eqno(4.6a)
$$
with leading term in asymptotic solution at
$r\to \infty $
as
$$
f (r)\sim r^{-1} (\exp (-(\sqrt{1 - e^2}r)/(2\sqrt{1 - e^2}) .
 \eqno(4.6b)
$$

Then the trial lowest eigenstate can be chosen as a product of these leading terms with yet
unknown energy.
Introducing notion
$\kappa=\sqrt{1-e^2}$ first we find the normalization condition
$$
 \int (\exp(-(\alpha/r) - r \kappa)/(4 \kappa))^2 r^2 dr=
 \frac{\left(\frac{\alpha }{\kappa }\right)^{3/2} K_3\left(4 \sqrt{\alpha  \kappa }\right)}{8 \alpha ^2 \kappa ^2}
 \eqno(4.7)
 $$
where
$K_i$ is the Bessel function of imaginary argument of order 
$i$.

Multiplying from the left the left-hand side of eq.() on normalized trial
function
$f(r)$ and integrate over radial coordinate.
As a result we get the expression of the form
$$
\sqrt{\frac{\alpha }{\kappa }} K_3\left(4 \sqrt{\alpha  \kappa
}\right) \left(\alpha  \kappa -\kappa ^2+\epsilon
^2-1\right)+\alpha  (2 \epsilon -1) K_2\left(4
   \sqrt{\alpha  \kappa }\right)=0
   \eqno(4.8)
$$
which is a second order algebraic equation on
$\epsilon$ with elementary solution for roots.

Now, in accord with usual Ritz variational method we have to minimize one of the root
on respect to $\kappa$ that leads to the approximate value of the lowerst eigenstate energy.
The appropriate dependence is shown in Fig.~9 for
$\alpha=1$.

\begin{figure*}[hbt]
\includegraphics[width=7cm]{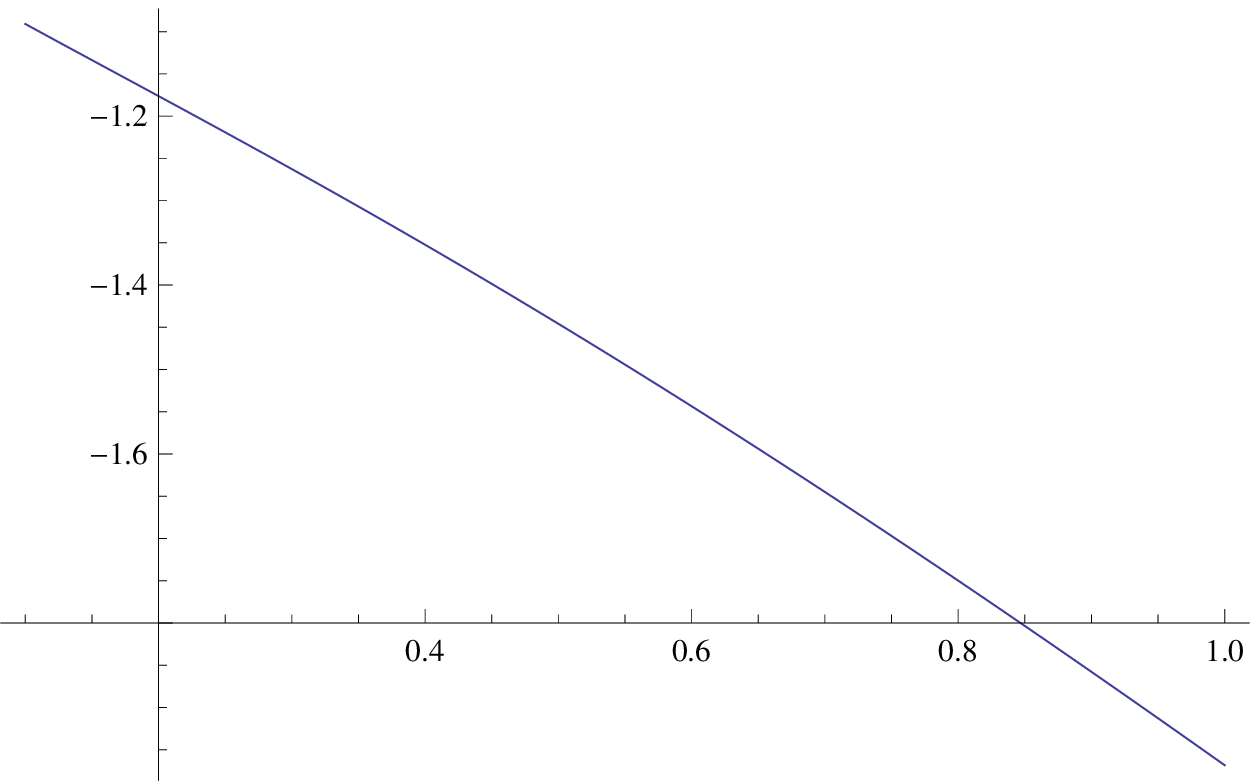} \hspace{0.5cm}
\includegraphics[width=7cm]{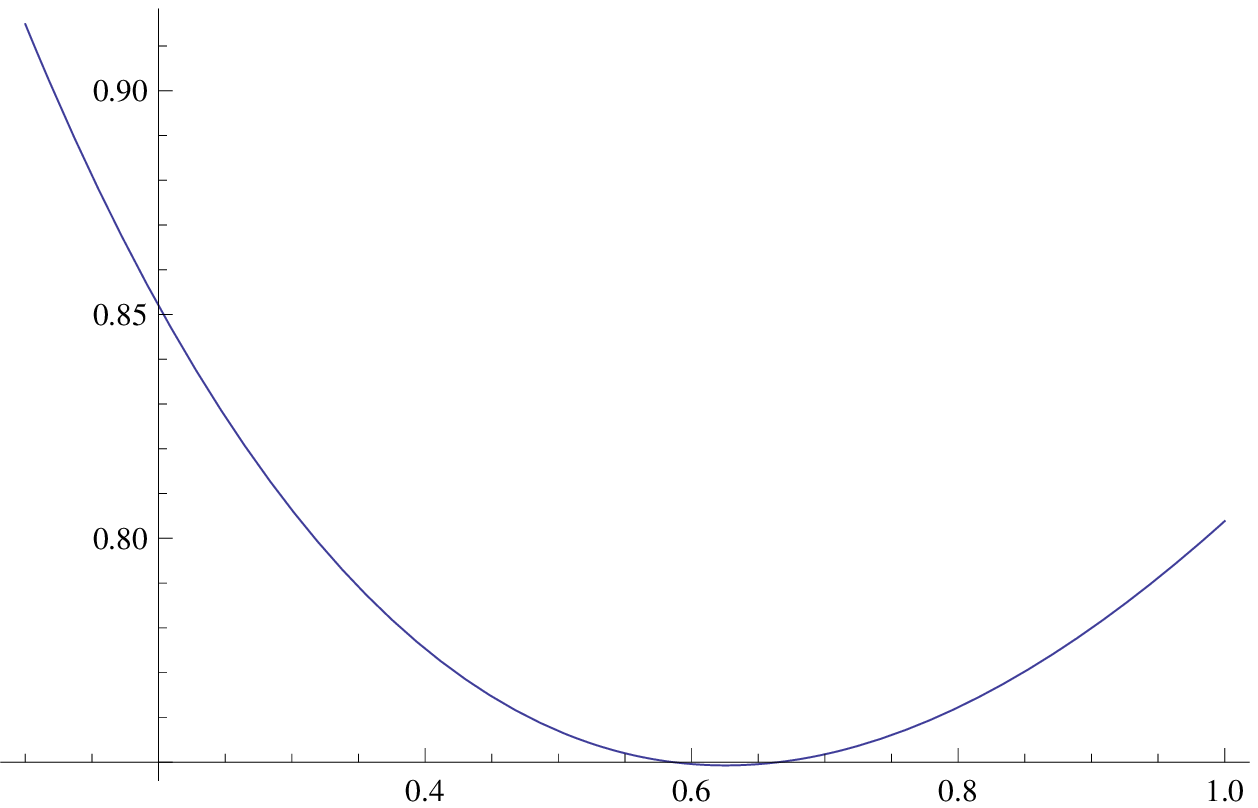}
 \caption {The dependencies of the roots $\epsilon_1(\kappa)$, $\epsilon_2(\kappa)$}
\label{fig4}
\end{figure*}

As one can see from Fig.~9 we indeed have a minimum for the second root,
its value  and location are
$e=0.749279$ at $\kappa=0.625342$.
The appropriate eigenfunction and squared redial momentum
$P^2(r)$  are shown in Fig.~10

\begin{figure*}[hbt]
\includegraphics[width=12cm]{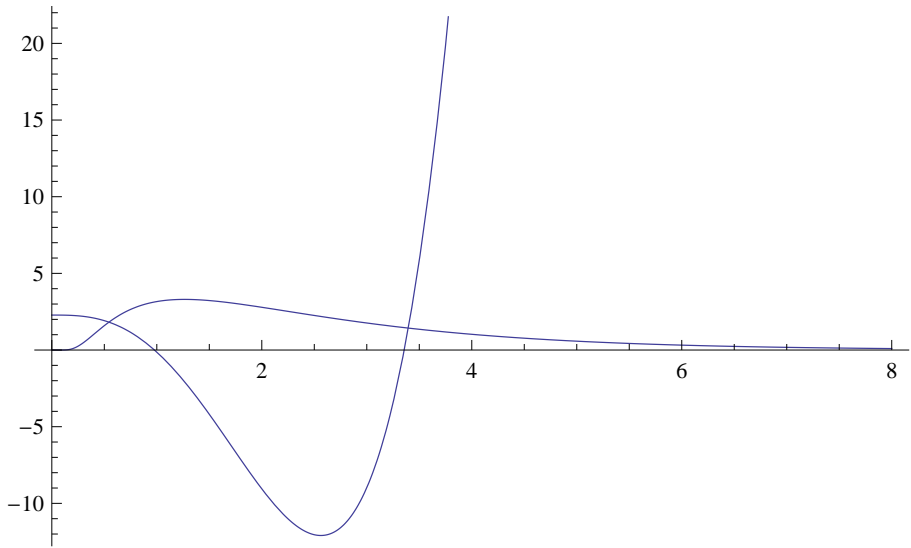}
 \caption {Plots of the lowest energy eigenfunction and $P^2(r)$ at $M=1, \sigma=-1, j=0,\alpha=1 $}
\label{fig4}
\end{figure*}

\end{document}